\documentclass[aps,prb,twocolumn,showpacs,preprintnumbers,amsmath,amssymb,superscriptaddress]{revtex4}

\usepackage{graphicx}% Include figure files
\usepackage{dcolumn}% Align table columns on decimal point
\usepackage{bm}% bold math

\begin{document}

\title{Dicke-like effect in spin-polarized transport through coupled quantum dots}

\author{Piotr Trocha}
\email{piotrtroch@gmail.com}\affiliation{Department of Physics,
Adam Mickiewicz University, 61-614 Pozna\'n, Poland}

\author{J\'ozef Barna\'s}
\email{barnas@amu.edu.pl} \affiliation{Department of Physics, Adam
Mickiewicz University, 61-614 Pozna\'n, Poland}
\affiliation{Institute of Molecular Physics, Polish Academy of
Sciences, 60-179 Pozna\'n, Poland}

\date{\today}
\begin{abstract}
Spin-dependent electronic transport through a quantum dot
side-coupled to two quantum dots and attached to ferromagnetic
leads with collinear (parallel and antiparallel) magnetizations is
analyzed theoretically. The intra-dot Coulomb correlations are
taken into account, whereas the inter-dot ones are neglected.
Transport characteristics, i.e. conductance and tunnel
magnetoresistance associated with the magnetization rotation from
parallel to antiparallel configurations, are calculated by the
noneqiulibrium Green function technique. The Green functions are
derived by the equation of motion method in the Hartree-Fock
approximation.  The conductance spectra are shown to reveal
features similar to the Dicke resonance in atomic physics.

\end{abstract}


\maketitle

\section{Introduction}

Interference effects in electronic transport through coupled
quantum dots (QDs) are of great interest from both fundamental and
application points of view. One of the simplest systems where the
interference effects are clearly visible, consists of two QDs
connected in parallel to electron reservoirs (metallic or
semiconducting leads). This interferometer-like geometry allows
observation of  such quantum phenomena like the Aharonov-Bohm
oscillations and the Fano antiresonance
\cite{guevara,lu,lu05,luliu,cho,chi,ding,johnson,trocha1,sztenkiel1,sztenkiel2,trocha2,tanaka}.

Conductance of systems consisting of three or more QDs reveals
further interesting features of quantum transport
\cite{emary,rogge,gaudreau,kuzmenko,saraga,zitko,guevara06,sun,aldea,gomez,wang}.
Of particular importance seems to be the interference phenomenon
\cite{brandes1,brandes2,brandes3,orellana1,orellana2}, which
resembles the well known Dicke resonance in atomic physics
\cite{dicke1,dicke2,devoe}. The key feature of the Dicke resonance
is the presence of a strong and very narrow emission line of a
collection of atoms which are separated by a distance smaller than
the wavelength of the emitted light \cite{dicke1,dicke2}.

In this paper we consider the Dicke-like resonance in spin
polarized transport through a system of three coupled QDs. One of
the QDs is attached to two ferromagnetic leads, while the other
two dots are side-coupled to the first one. Our considerations
include the intra-dot electron correlation, while the inter-dot
Coulomb interaction is assumed to be much smaller than the
intra-dot one and is omitted. Transport characteristics, in
particular conductance of the system and tunnel magnetoresistance
associated with rotation of the electrodes' magnetizations from
antiparallel to parallel alignment, are calculated by the
non-equilibrium Green function technique in the linear response
regime. Since the systems with Coulomb interactions cannot be
solved exactly, we applied the Hartree-Fock (HF) approximation
scheme to calculate the relevant Green functions from the
equations of motion.

In section 2 we describe the model, while theoretical formulation
of the problem and basic analytical formula are presented in
section 3. Numerical results for symmetric and asymmetric systems
are shown and discussed in section 4. It is shown there that
narrow Dicke-like peaks appear in the linear conductance. Final
conclusions are given in section 5.

\section{Description of the model}

In this paper we consider a system consisting of three
single-level quantum dots, coupled as shown schematically in
figure 1. The dots QD1 and QD3 are coupled to the dot QD2 {\it
via} direct hopping term. The dot QD2, in turn, is additionally
attached to ferromagnetic leads (see figure 1). Generally,
couplings between the dots as well as those between the dots and
external leads can be controlled by applied gate voltages. We also
note that technological difficulties associated with attaching a
semiconductor quantum dot to ferromagnetic metallic leads have
been overcome and first experimental results on spin polarized
transport through individual semiconductor quantum dots have been
published very recently \cite{hamaya}. Thus, experimental
investigation of spin polarized transport in systems of coupled
quantum dots seems to be only a matter of time. For simplicity we
consider only collinear (parallel and antiparallel) magnetic
orientations of the leads' magnetic moments. Apart from this, we
include the intra-dot Coulomb interactions, while the inter-dot
Coulomb repulsion is assumed to be weak and therefore is
neglected. The system under consideration is then described by
Hamiltonian of the general form $\hat{H}=\hat{H}_{\rm L}
+\hat{H}_{\rm R}+\hat{H}_{\rm QD}+\hat{H}_{\rm T}$. The terms
$\hat{H}_{\rm L}$ and $\hat{H}_{\rm R}$ describe respectively the
left and right electrodes in the non-interacting quasi-particle
approximation, $\hat{H}_{\alpha }=\sum_{{\mathbf k}\sigma}
\varepsilon_{{\mathbf k}\alpha \sigma}c^{\dagger}_{{\mathbf
k}\alpha \sigma} c_{{\mathbf k}\alpha \sigma}$ (for $\alpha ={\rm
L,R}$). Here, $c^{\dagger}_{{\mathbf k}\alpha \sigma}$
($c_{{\mathbf k}\alpha \sigma}$) is the creation (annihilation)
operator of an electron with the wave vector ${\mathbf k}$ and
spin $\sigma$ in the lead $\alpha$, whereas $\varepsilon_{{\mathbf
k}\alpha \sigma}$ denotes the corresponding single-particle
energy.

\begin{figure}
\begin{center}
\includegraphics[width=0.2\textwidth,angle=-90]{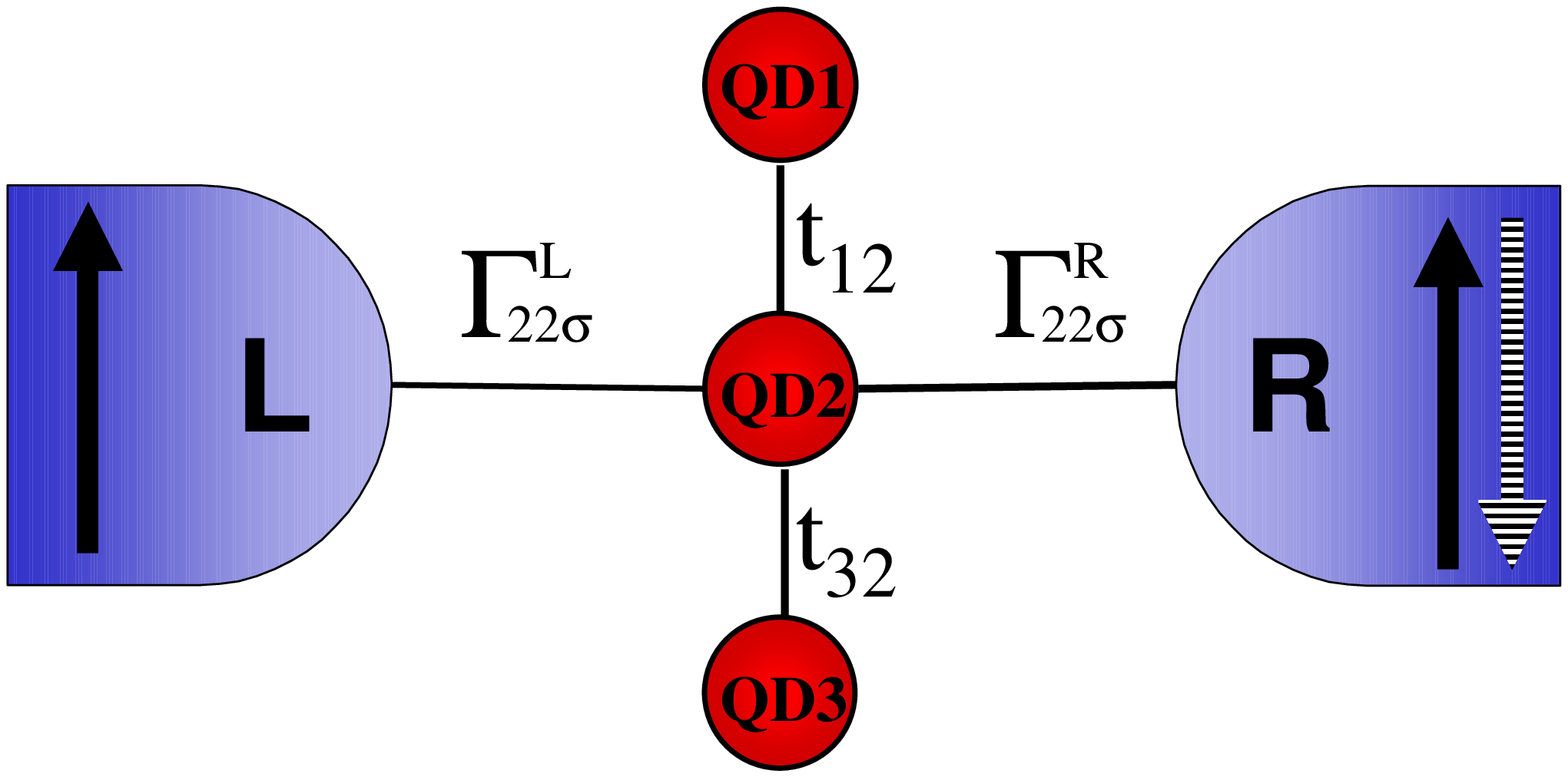}
\caption{Schematic picture of the quantum dot system coupled to
ferromagnetic leads.  $\Gamma_{22\sigma}^\alpha$ $(\alpha=L,R)$
describe spin dependent coupling of the dot QD2 to the
ferromagnetic leads, whereas $t_{j2\sigma}$ is the hopping
parameter between the $j$-th dot ($j=1,3$) and the dot QD2.}
\end{center}
\end{figure}

The third term of the Hamiltonian $H$ describes the system of
three coupled quantum dots,
   \begin{eqnarray}
   \hat{H}_{QD}=\sum_{i\sigma}\limits\epsilon_{i\sigma}d^\dag_{i\sigma}d_{i\sigma}
   +\sum_{i=1}^3\limits
   U_in_{i\sigma}n_{i\bar{\sigma}}
      \nonumber \\
   \qquad -\sum_{j(=1,3)}\sum_{\sigma}\limits(t_{j2\sigma}
  d^\dag_{2\sigma}d_{j\sigma}+h.c.),
   \end{eqnarray}
where $\epsilon_{i\sigma}$ is the energy level of the $i$-th dot
($i=1,2,3$), and $t_{j2\sigma}$ is the hopping parameter between
the $j$-th dot  ($j=1,3$) and the dot QD2. Both
$\epsilon_{i\sigma}$ and $t_{j2\sigma}$ are assumed to be spin
dependent in a general case. The second term of the Hamiltonian
$\hat{H}_{QD}$ describes the intra-dot Coulomb interactions, with
$U_i$ ($i=1,2,3$) denoting the corresponding Coulomb integrals.

The last term of the system's Hamiltonian, $H_{\rm T}$, describes
electron tunneling from the leads to the dot QD2 (or {\it vice
versa}), and takes the form
\begin{equation}
  \hat{H}_{\rm T}=\sum_{{\mathbf k}\alpha}\limits\sum_{\sigma}
   \limits (V_{2{\mathbf k}\sigma}^\alpha c^\dag_{{\mathbf k}\alpha\sigma}d_{2\sigma}+\rm h.c.)
   \end{equation}
where $V_{2{\mathbf k}\sigma}^\alpha$ are the relevant matrix
elements. Coupling of the dot to external leads can be
parameterized in terms of
$\Gamma^\alpha_{\sigma}(\epsilon)=2\pi\sum_{\mathbf k}
V_{2{\mathbf k}\sigma}^\alpha V^{\alpha\ast}_{2{\mathbf
k}\sigma}\delta(\epsilon-\epsilon_{{\mathbf k}\alpha\sigma})$. We
assume that $\Gamma^\alpha_{\sigma}(\epsilon )$ is constant within
the electron  band,
$\Gamma^\alpha_{\sigma}(\epsilon)=\Gamma^\alpha_{\sigma}={\rm
const}$ for $\epsilon\in\langle-D,D\rangle$, and
$\Gamma^\alpha_{\sigma}(\epsilon)=0$ otherwise. Here, $2D$ denotes
the electron band width. Introducing the spin polarization $p_{r}$
of lead $r$ ($r=L,R$) as $p_{r}=(\rho_{r}^{+}- \rho_{r}^{-})/
(\rho_{r}^{+}+ \rho_{r}^{-})$, the coupling parameters can be
expressed as $\Gamma_{r}^{+(-)}=\Gamma_{r}(1\pm p_{r})$, with
$\Gamma_{r}= (\Gamma_{r}^{+} +\Gamma_{r}^{-})/2$. Here,
$\rho_{r}^{+}$ and $\rho_{r}^{-}$ are the densities of states at
the Fermi level for spin-majority and spin-minority electrons in
the lead $r$, while $\Gamma_{r}^{+}$ and $\Gamma_{r}^{-}$ describe
coupling of the dot QD2 to the lead $r$ in the spin-majority and
spin-minority channels, respectively.

\section{Analytical description}

Electric current $J$ flowing through the system is given by  the
formula \cite{meir92,jauho}
  \begin{eqnarray}
   J=\frac{ie}{2\hbar}\int\frac{d\epsilon}{2\pi}
  Tr\{[\mathbf{\Gamma}^L(\epsilon)-\mathbf{\Gamma}^R(\epsilon)]\mathbf{G}^<(\epsilon)
     \nonumber \\
  \quad + [f_L(\epsilon)\mathbf{\Gamma}^L(\epsilon)-
  f_R(\epsilon)\mathbf{\Gamma}^R(\epsilon)][\mathbf{G}^r(\epsilon)-\mathbf{G}^a(\epsilon)]\},
  \end{eqnarray}
where,
$f_\alpha(\epsilon)=[e^{(\epsilon-\mu_\alpha)/k_BT}+1]^{-1}$ is
the Fermi-Dirac distribution function in the lead $\alpha$,
$\mathbf{G}^<(\epsilon)$ and $\mathbf{G}^{r(a)}(\epsilon)$ are the
Fourier transforms of the lesser and retarded (advanced) Green
functions of the dots, and $\mathbf{\Gamma}^\alpha(\epsilon)$ (for
$\alpha=L,R$) is a matrix which describes coupling of the dots to
the leads. In the case under consideration the matrix
$\mathbf{\Gamma}^\alpha(\epsilon)$ takes a simple form with only
one nonzero matrix element,
   \begin{equation}\label{}
  \mathbf{\Gamma}^\alpha_\sigma=\left(%
\begin{array}{ccc}
  0 & 0 & 0 \\
  0 & \Gamma^\alpha_{22\sigma} & 0 \\
  0 & 0 & 0
\end{array}
\right).
\end{equation}
In the case under consideration one can write
$\Gamma^L_{22\sigma}=\Gamma(1\pm p_L)$ for coupling of the dot to
the left lead, and $\Gamma^R_{22\sigma}=\gamma\Gamma(1\pm p_R)$
(parallel configuration) or $\Gamma^R_{22\sigma}=\gamma\Gamma(1\mp
p_R)$ (antiparallel configuration) for coupling to the right lead.
Here, $\sigma=\uparrow$ (upper sign), $\sigma=\downarrow$ (lower
sign), $\Gamma$ is a constant, whereas $\gamma$ describes
asymmetry in the coupling of the dot to the left and right leads.

To calculate electric current one needs to know the Green
functions $G^{r(a)}_{ij\sigma}(\epsilon)$. These can be found by
the equation of motion method. First, we apply the equation of
motion to the causal Green function $G_{ij\sigma}(\epsilon)$. This
equation of motion generates higher order Green functions, for
which we write the corresponding equations of motion. New higher
order Green functions are then decoupled in terms of the
Hartree-Fock decoupling scheme. This decoupling scheme, described
in details in Refs \cite{lu,trocha2}, allows to express the
higher-order Green functions through the lower-order ones. This,
in turn, allows closing the relevant system of equations for the
Green function $G_{ij\sigma}(\epsilon)$ and to write it in the
matrix form as
\[    \left(
\begin{array}{ccc}
  \epsilon-\epsilon_{1\sigma}-A_{1} & t_{12\sigma}B_{1} & 0 \\
  t_{12\sigma}B_{2} & \epsilon-\epsilon_{2\sigma}-\Sigma_\sigma-A_{2} & t_{32\sigma}B_{2} \\
  0 & t_{32\sigma}B_{3} & \epsilon-\epsilon_{3\sigma}-A_{3} \\
\end{array}
\right)
\]
\begin{equation}\label{}
\times \left(
\begin{array}{c}
  G_{1j\sigma} \\
  G_{2j\sigma} \\
  G_{3j\sigma} \\
\end{array}
\right)=\left(
\begin{array}{c}
  \delta_{1j}B_{1} \\
  \delta_{2j}B_{2} \\
  \delta_{3j}B_{3} \\
\end{array}
\right),
\end{equation}
where for simplicity the $\epsilon$-dependence of the Green
functions and other parameters has not been indicated explicitly
(it will be restored where necessary). Apart from this, we have
introduced the following notation:
\begin{equation}
A_i=\frac{U_i(n_{i2\bar{\sigma}}-n_{2i\bar{\sigma}})}
{\epsilon-\epsilon_{i\sigma}-U_i}
\end{equation}
for $i=1,3$,
\begin{eqnarray}
A_2=U_2(\epsilon-\epsilon_{2\sigma}-U_2)^{-1}
[t_{12\sigma}(n_{12\bar{\sigma}}-n_{21\bar{\sigma}}) \nonumber \\
+ t_{32\sigma}(n_{32\bar{\sigma}}-n_{23\bar{\sigma}})
+(C^\dag_{2\bar{\sigma}}-C_{2\bar{\sigma}})
n_{2\bar{\sigma}}\Sigma_{\sigma}],
\end{eqnarray}
and
\begin{equation}
B_i=1+\frac{U_in_{2\bar{\sigma}}}{\epsilon-\epsilon_{i\sigma}-U_i}
\end{equation}
for $i=1, 2, 3$. To simplify notation we put here $\langle
n_{ij\sigma}\rangle = \langle d^+_id_j\rangle =n_{ij\sigma}$,
$\langle n_{i\sigma}\rangle = \langle d^+_id_i\rangle
=n_{i\sigma}$, and
$C_{2\bar{\sigma}}\equiv\sum_{k\alpha}V^\alpha_{2k\bar{\sigma}}\langle
c^\dag_{k\alpha\bar{\sigma}}d_{2\bar{\sigma}}\rangle$. We have
also defined the self-energy $\Sigma_{\sigma}$ as
\begin{equation}\label{}
    \Sigma_{\sigma}=\Sigma^L_{\sigma}+\Sigma^R_{\sigma},
\end{equation}
with
\begin{equation}\label{}
    \Sigma^\alpha_{\sigma}=\sum_k\limits\frac{V_{2k\sigma}^\alpha V_{2k\sigma}^{\alpha\ast}}
    {\epsilon-\epsilon_{k\alpha\sigma}}
\end{equation}
for $\alpha =L,R$.

Having found the causal Green functions from equation 5, one can
calculate the retarded (advanced) Green functions as
$G^{r(a)}_{ij\sigma}(\epsilon)=G_{ij\sigma}(\epsilon\pm i0^+)$.
The corresponding self-energy
$\Sigma^{r(a)}_{\sigma}(\epsilon)=\Sigma_{\sigma}(\epsilon\pm
i0^+)$ takes then the form,
\begin{equation}\label{}
    \Sigma^{r(a)}_{\sigma}(\epsilon)=\Lambda_{22\sigma}(\epsilon)\mp
    \frac{i}{2}\Gamma_{22\sigma},
\end{equation}
with
\begin{equation}\label{}
    \Gamma_{22\sigma}=\Gamma^L_{22\sigma}+\Gamma^R_{22\sigma},
\end{equation}
and
\begin{equation}\label{}
    \Lambda_{22\sigma}(\epsilon)=
    -\frac{1}{2\pi}\Gamma_{22\sigma}\ln\left(\frac{D-\epsilon}{D+\epsilon}\right).
\end{equation}

Now, we need to find the lesser Green function. This can be found
from the corresponding equation of motion, with the higher order
Green functions calculated on taking into account the Langreth
theorem \cite{jauho} and the Hartree-Fock decoupling scheme used
when calculating the causal Green function.

\section{Numerical results}

Using the formulas derived above, we calculate numerically the
basic transport characteristics, i.e. the conductance and tunnel
magnetoresistance. The latter quantity describes the resistance
change when magnetic configuration of the system varies from
antiparallel to parallel, and is described quantitatively by the
ratio $(R_{\rm AP}-R_{\rm P})/R_{\rm P}$, where $R_{\rm AP}$ and
$R_{\rm P}$ denote the resistance  in the antiparallel and
parallel magnetic configurations, respectively. We assume that the
dot levels are spin degenerate, $\epsilon_{i\sigma}=\epsilon_i$
(for $i =1,2,3$), and introduce two parameters, $h_1$ and $h_2$,
which describe separation of the levels $\epsilon_1$ and
$\epsilon_3$ from the level $\epsilon_2$, i.e.
$\epsilon_{1}=\epsilon_2-h_1$ and $\epsilon_{3}=\epsilon_2+h_2$.
Moreover, we assume $p_L=p_R=p$ and $\gamma =1$.

\begin{figure}
\begin{center}
\includegraphics[width=0.48\textwidth,angle=0]{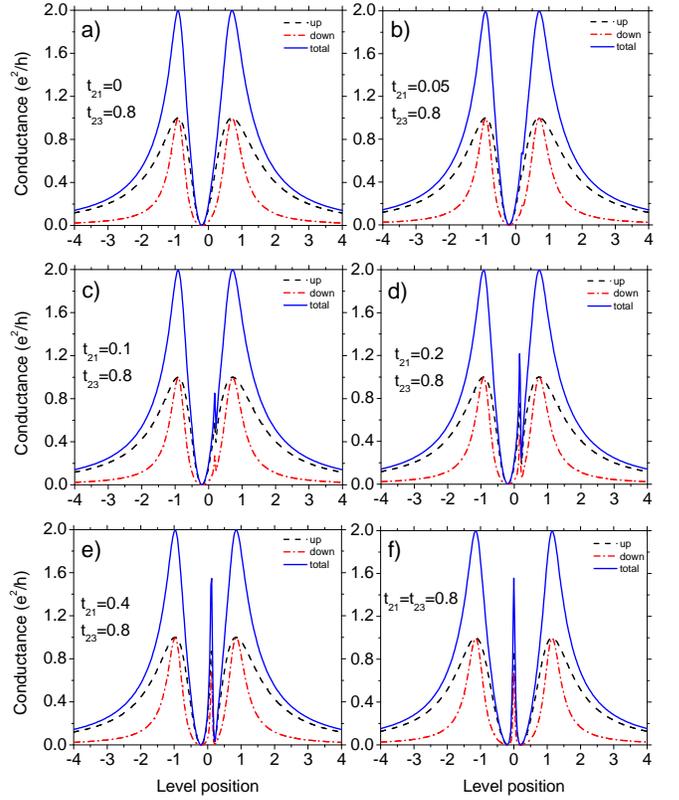}
\caption{Linear conductance in the parallel magnetic
configuration, calculated as a function of the energy level
$\epsilon_2$ of QD2 for indicated values of the intra-dot coupling
parameters $t_{21}$ and $t_{23}$ and for $p=0.4$, $U_1=U_2=U_3=0$,
$h_1=h_2=0.2$, $\gamma =1$,  and $k_BT=0.01$. The energy is
measured in the units of $\Gamma$.}
\end{center}
\end{figure}

Consider first the limit of vanishing Coulomb interaction, $U_i=0$
for $i=1,2,3$. In figure 2 we show the linear conductance in the
parallel magnetic configuration, calculated as a function of the
energy level $\epsilon_2$ of QD2, measured from the Fermi level of
the leads. The total conductance as well as that of individual
spin channel are shown there. The energy level of the dot QD1 is
shifted down by 0.2 with respect to $\epsilon_2$, while that of
QD3 is shifted up by the same amount. Here and in the following
the energy is measured in the units of $\Gamma$ ($\Gamma
=\Gamma_L=\Gamma_R$). The hoping parameter $t_{23}$ is constant
while the parameter $t_{21}$ increases from zero in figure 2(a) to
$t_{21}=t_{23}$ in figure 2(f). Let us analyze now the main
features of the conductance spectra. In figure 2(a) the dot QD1 is
entirely detached from the dot QD2 and therefore has no influence
on transport. The other two dots are coupled and form bonding and
anti-bonding like states. Due to a nonzero value of $h_2$, these
states are coupled differently to the leads. The linear
conductance shows then two relatively broad peaks related to the
bonding and anti-bonding like states, with a deep minimum in
between, where the conductance turns to zero due to destructive
interference of electron waves transmitted via bonding and
anti-bonding states. We note, that the broadening of peaks is
determined mainly by the coupling to the leads, but temperature
also plays a role.

When the dot QD1 becomes weakly coupled to the dot QD2, a narrow
peak emerges in the conductance at the position corresponding to
the energy level of the dot QD1 (see figure 2(b-d)). The onset of
this peak is a direct consequence of a nonzero value of $t_{21}$.
The dots QD1 and QD3 are then both coupled to the dot QD2 and,
disregarding coupling to the leads, one finds three new energy
level of the whole three-dot system. The level corresponding to
the new peak is only weakly coupled to the leads. When $t_{21}$
increases further, this peak moves towards zero energy, and in the
strictly symmetrical case, $t_{21}=t_{23}$, the peak appears
strictly at the position corresponding to the bare level of the
dot QD2, $\epsilon_2=0$. The spectrum becomes then symmetrical,
which is a consequence of the symmetry in the system. The central
peak, however, remains much narrower than the other two peaks.
Similar scenario also holds for constant $t_{21}$ and $t_{23}$
increasing from zero to $t_{23}=t_{21}$. The new peak emerges now
on the left side of the minimum and moves towards $\epsilon_2=0$,
as before. All the features discussed above appear not only in the
total conductance, but also in the conductance of each spin
channel separately, as clearly seen in figure 2. The heights of
the side peaks in figure 2 are equal to 2 conductance quanta
($e^2/h$), independently of the values of the inter-dot hoping
parameters. The height of central narrow peak, however, is smaller
than that of the side broad peaks.

\begin{figure}
\begin{center}
\includegraphics[width=0.48\textwidth,angle=0]{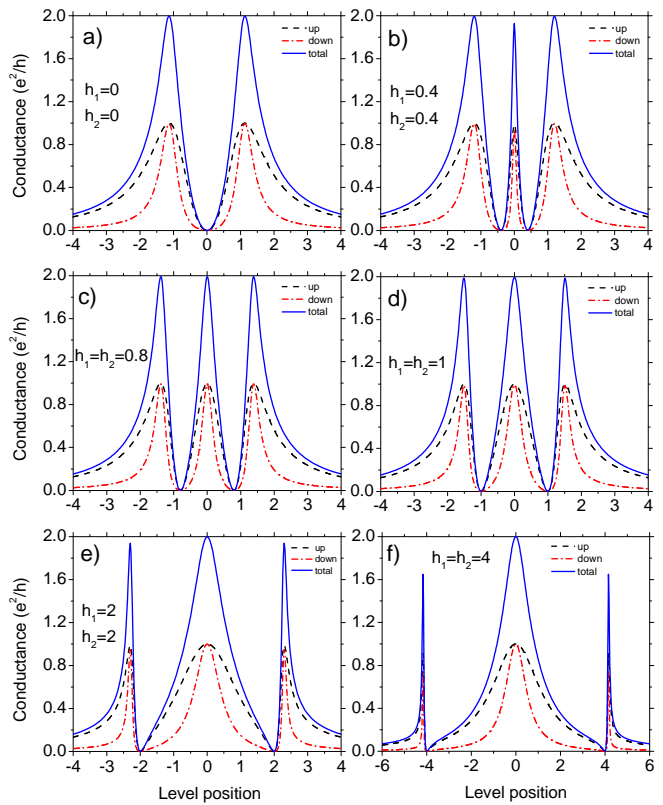}
\caption{Linear conductance in the parallel magnetic
configuration, calculated as a function of the energy level
$\epsilon_2$ of QD2 for indicated values of the level shifts
 $h_{1}$ and $h_{2}$, and for $t_{21}=t_{23}=0.8$,
$p=0.4$, $\gamma =1$, $U_1=U_2=U_3=0$, and $k_BT=0.01$. The energy
is measured in the units of $\Gamma$.}
\end{center}
\end{figure}

The central narrow peak resembles the Dicke resonance in atomic
physics, where a strong narrow emission line appears when the
distance between atoms is smaller than the Fermi wavelength of the
corresponding radiation. In our case the role of distance between
atoms is plied by the distance in energy levels, $h_1$ and $h_2$.
When $h_1$ and $h_2$ increase, the width of the central line also
increases. The two side peaks, in turn, become narrower and
narrower. This is shown explicitly in figure 3, where the linear
conductance for the symmetrical situation is presented for
increasing $h_1$ and $h_2$ (the other parameters are the same as
in figure 2). This behavior is quite reasonable, as the
eigen-states of the dot system become more and more localized on
the corresponding individual dots with increasing $h_1$ and $h_2$.
Since the dots QD1 and QD3 are not coupled directly to the leads,
the corresponding side peaks in the conductance become narrower
and their width decreases with increasing $h_1$ and $h_2$. The
central zero-energy level becomes then localized on the dot QD2,
and the corresponding peak in the conductance becomes broader.
Moreover, height of the central peaks increases and becomes equal
to two conductance quanta, while that of the side peaks decreases
with increasing $h_1$ and $h_2$.

When $h_1=h_2\neq 0$ and $t_{12}=t_{23}=t\neq 0$ (as in figure 3),
the two side peaks in the conductance appear at the positions
$\pm\sqrt{h^2+2t^2}$. The third (central) peak occurs at
$\epsilon=0$, while the conductance reaches zero at $\pm h$. A
particular situation occurs when $h_1=h_2=0$. The central peak,
clearly seen in figure 3 for $h_1,h_2\neq 0$, does not occur then
in the conductance. Since the system of three single-level QD's
has three molecular-like states (denoted with the index 1, 2 and 3
for increasing energy), one could expect also three peaks in the
conductance. However, the matrix elements of coupling between the
molecular state $|\tilde{2}\rangle$ and the left and right leads
vanish, the molecular state $|\tilde{2}\rangle$ becomes decoupled
from the leads for $h_1=h_2=0$, which leads to the absence of the
central peak in the conductance. As a consequence, one observes
only two peaks in the linear conductance, situated at the
positions $\epsilon=\pm\sqrt{2}t$, and the transmission reaches
zero at $\epsilon=0$ (eigenvalue of the state
$|\tilde{2}\rangle$). To show this more formally, let us perform
the transformation of the dot's operators
\begin{equation}
\tilde{d}_{1\sigma}=\frac{1}{2}(d_{1\sigma}+\sqrt{2}d_{2\sigma}+d_{3\sigma}),
\end{equation}
\begin{equation}
\tilde{d}_{2\sigma}=\frac{1}{\sqrt{2}}(d_{1\sigma}-d_{3\sigma}),
\end{equation}
\begin{equation}
\tilde{d}_{3\sigma}=\frac{1}{2}(d_{1\sigma}-\sqrt{2}d_{2\sigma}+d_{3\sigma}).
\end{equation}
Assuming $\epsilon_{i\sigma}=\epsilon_{0}$ for $i=1,2,3$, the
Hamiltonian of the isolated triple dots becomes then diagonal and
takes the form
\begin{equation}
\tilde{H}_{QD}=(\epsilon_0+\sqrt{2}t)\tilde{d}^\dag_{1\sigma}\tilde{d}_{1\sigma}+
\epsilon_0\tilde{d}^\dag_{2\sigma}\tilde{d}_{2\sigma}+
(\epsilon_0-\sqrt{2}t)\tilde{d}^\dag_{3\sigma}\tilde{d}_{3\sigma}.
\end{equation}
In turn, the tunnelling Hamiltonian acquires then the form
\begin{equation}
\tilde{H}_{T}=\sum_{{\mathbf k}\alpha\sigma}\limits\sum_{i=1,3}
   \limits(\tilde{V}_{i{\mathbf k}\sigma}^\alpha c^\dag_{{\mathbf k}\alpha\sigma}
   \tilde{d}_{i\sigma}+\rm h.c.),
\end{equation}
   where
\begin{equation}
\tilde{V}_{1{\mathbf
k}\sigma}^\alpha=\frac{\sqrt{2}}{2}V_{2{\mathbf k}\sigma}^\alpha ,
\end{equation}
\begin{equation}
\tilde{V}_{3{\mathbf k}\sigma}^\alpha=-\frac{\sqrt{2}}
{2}V_{2{\mathbf k}\sigma}^\alpha ,
\end{equation}
and

\begin{equation}
\tilde{V}_{2{\mathbf k}\sigma}^\alpha=0.
\end{equation}
This clearly shows that the central molecular-like level becomes
effectively decoupled from the leads.

\begin{figure}
\begin{center}
\includegraphics[width=0.48\textwidth,angle=0]{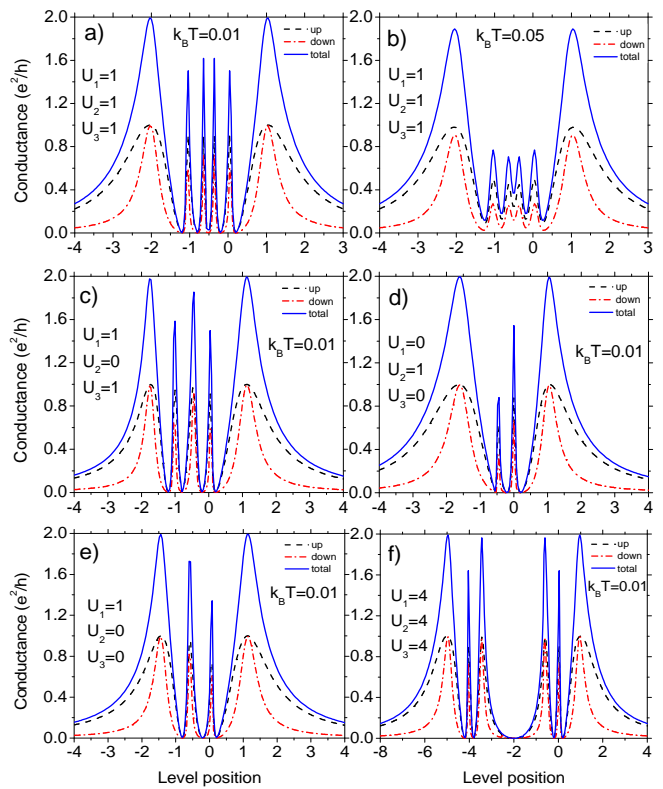}
\caption{Linear conductance in the parallel magnetic
configuration, calculated as a function of the energy level
$\epsilon_2$ of QD2 for indicated values of temperature and
intra-dot Coulomb repulsion parameters $U_1,U_2,U_3$. The other
parameters are:  $t_{21}=t_{23}=0.8$, $h_1=h_2=0.2$, $p=0.4$, and
$\gamma =1$. The energy is measured in the units of $\Gamma$.}
\end{center}
\end{figure}

Let us consider now the influence of Coulomb correlation in the
dots (nonzero parameters $U_i$ for $i=1,2,3$) and nonzero
temperature  on the spectra described above. Figure 4(a)
corresponds to figure 2(f), except the Coulomb parameters which in
figure 4(a) are nonzero, $U_1=U_2=U_3=1$, while in figure 2(f)
they vanish, $U_1=U_2=U_3=0$. Owing to the nonzero Coulomb
parameters, the number of peaks in the conductance becomes doubled
in comparison to that in the case with vanishing Coulomb
interaction. The side peaks are broad as in figure 2(f), whereas
all the other four peaks are very narrow. Figure 4(b) shows the
conductance spectrum for the same parameters as in figure 4(a),
except temperature which now is higher. The peak heights are now
smaller and the peaks become broader, as one might expect. Figures
4(c-e) show the conductance for a system with the Coulomb
parameter vanishing for one or two dots. When the Coulomb
parameter of a particular dot vanishes, the number of peaks
decreases by one. In all the cases, however, the side peaks are
broad while the others are narrow. Figure 4(f) corresponds to
figure 4(a), but calculated for larger Coulomb correlation
parameters, $U_1=U_2=U_3=4$. The whole spectrum becomes clearly
split into two parts; one is similar to that for vanishing Coulomb
parameters, and the second is its Coulomb counterpart, with a
mirror plane somewhere in between the two parts.

\begin{figure}
\begin{center}
\includegraphics[width=0.38\textwidth,angle=0]{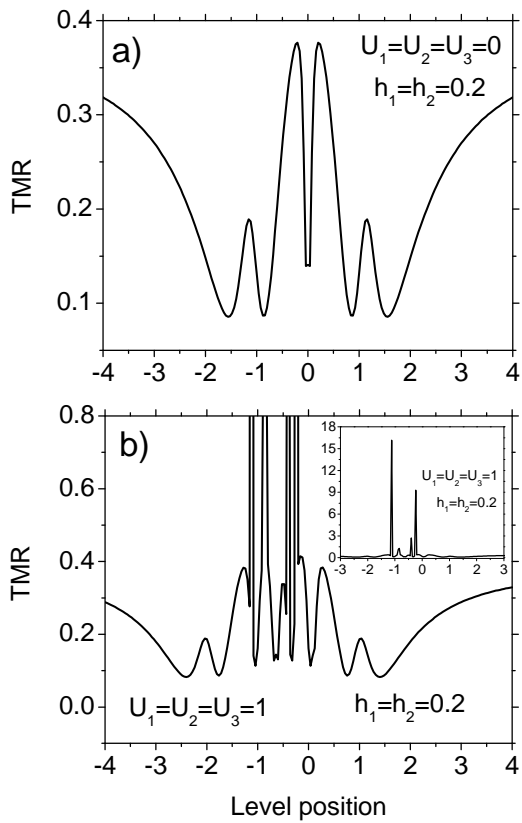}
\caption{TMR in the linear response regime, calculated as a
function of the energy level $\epsilon_2$ of QD2 for indicated
values of the intra-dot Coulomb repulsion parameters
$U_1,U_2,U_3$, and for $t_{21}=t_{23}=0.8$, $h_1=h_2=0.2$,
$p=0.4$, $\gamma  =1$, and $k_BT=0.01$. The energy is measured in
the units of $\Gamma$.}
\end{center}
\end{figure}

All the results presented above were calculated for the parallel
configuration of the leads' magnetic moments. Qualitatively
similar results have been obtained for the antiparallel magnetic
configuration. The quantitative difference leads to a nonzero TMR.
In figure 5 we show TMR in the linear response regime as a
function of the energy level $\epsilon_2$ of QD2. When the dots'
levels are well above or well below the Fermi level of the leads,
the TMR tends to the Julliere's value observed in planar magnetic
tunnel junctions. When, in turn, the dot levels cross the Fermi
energy, the situation becomes more complex and TMR significantly
depends on the Coulomb parameters $U_i$. In the limit of $U_i=0$
for $i=1,2,3$, see figure 5(a), TMR shows then a clear peak
structure and fluctuates about the Julliere's value. Qualitatively
similar behavior of TMR was also found in transport through single
quantum dots \cite{Weymann}. Variation of TMR with the level
position is more complex for nonzero values of the Coulomb
correlation parameters, as shown in figure 5(b). Significant
resonance-like peaks in the TMR occur now at certain values of the
level positions, at which the conductance (current) is strongly
suppressed by the interference effects. Presence of such peaks in
TMR is a consequence of the fact that the conditions for full
destructive interference depend on magnetic configuration of the
system for nonzero values of $U_i$. Outside the resonance-like
peaks the behavior of TMR is similar to that for vanishing $U_i$,
except that the number of different peaks is larger due to the
Coulomb counterparts in the conductance spectra.

\section{Conclusions}
In conclusion, we have calculated conductance of three coupled
quantum dots. The conductance spectra, calculated in the
Hartree-Fock approximation, clearly show presence of a central
narrow peak, which resemblances the Dicke resonance in optical
emission spectra of atoms. We have also shown, that the Coulomb
correlations on the dots lead to doubling of the resonance peaks
in the conductance. Additionally, we have calculated TMR effect
associated with leads' magnetic moments rotation from antiparallel
to parallel alignment, and shown that the intra-dot Coulomb
correlation leads to some resonance-like enhancement of TMR at
some values of the level positions.

\begin{acknowledgements}
 This work, as part of the European Science Foundation EUROCORES
Programme SPINTRA, was supported by funds from the Ministry of
Science and Higher Education as a research project in years
2006-2009 and the EC Sixth Framework Programme, under Contract N.
ERAS-CT-2003-980409.
\end{acknowledgements}

\maketitle

\end{document}